\documentclass[nobibnotes,superscriptaddress,showpacs,preprintnumbers,twocolumn,prd]{revtex4-1}

\usepackage{amsmath,amssymb}

\begin{document}

\title{Constraints on leptogenesis from a symmetry viewpoint}

%\date{\today}

\author{R.~Gonz\'{a}lez Felipe}
\email{gonzalez@cftp.ist.utl.pt}
\affiliation{Area Cient\'{\i}fica de F\'{\i}sica, Instituto Superior de Engenharia de Lisboa, Rua Conselheiro Em\'{\i}dio Navarro 1, 1959-007 Lisboa, Portugal}
\affiliation{Departamento de F\'{\i}sica and Centro de F\'{\i}sica Te\'orica de Part\'{\i}culas (CFTP), Instituto Superior T\'ecnico, Av. Rovisco Pais, 1049-001 Lisboa, Portugal}

\author{H.~Ser\^odio}
\email{hserodio@cftp.ist.utl.pt}
\affiliation{Departamento de F\'{\i}sica and Centro de F\'{\i}sica Te\'orica de Part\'{\i}culas (CFTP), Instituto Superior T\'ecnico, Av. Rovisco Pais, 1049-001 Lisboa, Portugal}

\begin{abstract}
It is shown that type I seesaw models based on the standard model Lagrangian extended with three heavy Majorana right-handed fields do not have leptogenesis in leading order, if the symmetries of mass matrices are also the residual symmetry of the Lagrangian. In particular, flavor models that lead to a mass-independent leptonic mixing have a vanishing leptogenesis \textit{CP} asymmetry. Based on symmetry arguments, we prove that in these models the Dirac-neutrino Yukawa coupling combinations relevant for leptogenesis are diagonal in the physical basis where the charged leptons and heavy Majorana neutrinos are diagonal.
\end{abstract}
\maketitle

\section{Introduction}

After Harrison, Perkin, and Scott (HPS)~\cite{Harrison:2002er} pointed out that leptonic mixing at low energies could be described by the so-called tribimaximal mixing, many attempts at explaining this pattern using continuous or discrete symmetry groups have been prompted. Besides specifying the mass matrix textures, any flavor symmetry imposed to a model also leads to a particular mixing pattern. Furthermore, textures coming from symmetries are usually very restrictive, since not only do they forbid some couplings, but also yield relations among them. Among these textures, the so-called mass-independent textures~\cite{Lam:2006wm}, i.e. textures where the mass matrix diagonalization is independent of mass parameters, are common in models with flavor symmetries. In the leptonic sector, the advantage of assuming such a setup is that the Pontecorvo-Maki-Nakagawa-Sakata (PMNS) leptonic mixing matrix turns out to be independent of any mass parameter or hierarchy, without any accidental cancellation or fine tuning. In particular, the HPS tribimaximal mixing can be obtained from a $\mu-\tau$ and magic symmetric low-energy neutrino mass matrix~\cite{Lam:2006wm}. Notice however that tribimaximal mixing does not necessarily require this setup. Rather than a consequence of an underlying flavor symmetry, it could arrive from various low-energy constructions~\cite{Chan:2007ng}.

By predicting the leptonic mixing and neutrino mass spectrum, models based on flavor symmetries also offer the possibility to address other fundamental problems such as the matter-antimatter asymmetry observed in the Universe. Among several viable mechanisms, leptogenesis~\cite{Fukugita:1986hr} has become one of the most attractive scenarios. In its simplest realization, the standard model (SM) is extended to include heavy singlet right-handed neutrinos (type I) or heavy scalar triplets (type II), which also provide a natural explanation for the tiny masses of the light neutrinos through the seesaw mechanism~\cite{seesaw}.

Recently, leptogenesis from type I seesaw and flavor symmetries with exact HPS mixing have been put into conflict~\cite{Jenkins:2008rb}. In particular, it has been argued that the models in the literature that generate an exactly tribimaximal leptonic mixing matrix using a flavor symmetry do not have leptogenesis and, consequently, higher-order symmetry breaking corrections are required in these models to produce a nonvanishing leptogenesis asymmetry.  Nevertheless, in Ref.~\cite{Branco:2009by}, a model was proposed which leads to the HPS leptonic mixing pattern and, simultaneously, to an exact degeneracy of the heavy Majorana neutrinos. Although in this framework there is no leptogenesis in leading order, it becomes viable if the right-handed neutrino degeneracy is lifted either by renormalization group effects or by a soft breaking of the flavor symmetry, without the need of higher-order corrections.

One may wonder whether the above feature, namely, the absence of leptogenesis in models with flavor symmetries and HPS leptonic mixing, is specific to these models or it is a more general property. The aim of this work is to demonstrate that the type I seesaw models commonly discussed in the literature and, particularly, flavor models that lead to a mass-independent leptonic mixing, have a vanishing leptogenesis \textit{CP} asymmetry in leading order. In these models, the relevant Dirac-neutrino mass matrix combinations are real and in general diagonal, thus preventing leptogenesis to occur. If a hybrid type I/II seesaw framework is invoked, leptogenesis can then be realized in a more natural way.

\section{Symmetries of Majorana mass matrices}
Majorana mass terms of the form
\begin{align}\label{efflag}
\mathcal{L_\text{eff}}=\frac{1}{2}\overline{\nu_L}m_\nu\nu_L^c+\text{H.c.}\,,
\end{align}
are common in most of the SM extensions. Mass matrices arising from these terms are symmetric and, in general, complex. Any symmetric matrix and, in particular, the $3\times3$ effective Majorana mass matrix for light neutrinos $m_\nu$, has the symmetry
\begin{align}\label{GL}
\mathcal{G}_L^\dagger m_\nu\mathcal{G}_L^\ast=m_\nu\,,
\end{align}
where $\mathcal{G}_L$ is a unitary matrix. The matrix $m_\nu$ can be diagonalized through a unitary matrix $U_\nu$, such that
\begin{align}\label{diagm}
U^\dagger_\nu m_\nu U_\nu^\ast=d_\nu\,,
\end{align}
where $d_\nu=\text{diag}(m_1,m_2,m_3)$, with $m_i$ real and positive. From Eqs.~(\ref{GL}) and (\ref{diagm}) one obtains
\begin{align}\label{GLp}
\mathcal{G}^{\prime \dagger}_Ld_\nu\mathcal{G}^{\prime \ast}_L=d_\nu\,,
\end{align}
where
\begin{equation}\label{GLcons}
\mathcal{G}^\prime_L = U_\nu^\dagger \mathcal{G}_L U_\nu\,.
\end{equation}
Clearly, the generators $\mathcal{G}_L$, under which the effective neutrino mass matrix
$m_\nu$ is invariant according to Eq.~(\ref{GL}), are built from the columns of the matrix $U_\nu$ that diagonalizes $m_\nu$.

Let us find the possible choices for the symmetry groups under which the matrices $\mathcal{G}_{L}^\prime$ are invariant. It is clear that after the diagonalization of the Majorana mass matrix $m_\nu$, there is always a freedom to redefine the Majorana fields $\nu_{Li} \rightarrow \pm\, \nu_{Li}\,$.  Obviously, this transformation corresponds to the $Z_2\times Z_2\times Z_2$ symmetry group and leaves $d_\nu$ diagonal, real and positive. If $\mathcal{G}^\prime_L$ belongs to $SU(3)$ then the symmetry is reduced to $Z_2\times Z_2$. Note also that the symmetry group of $\mathcal{G}_L^\prime$ is also the symmetry group of $\mathcal{G}_L$, since they are connected by a similarity transformation.

Now we may ask the following question: Is $Z_2\times Z_2\times Z_2$ the maximal symmetry group of the Majorana mass matrix $m_\nu$ or can one have another symmetry group $G$, such that $G \supset Z_2\times Z_2\times Z_2$? In order to answer this question, let us consider Eq.~(\ref{GLp}), rewritten as
\begin{align}
\mathcal{G}_L^{\prime \dagger}d_\nu=d_\nu \mathcal{G}_L^{\prime T}\,.
\end{align}
This equation leads to relations of the type
\begin{align}\label{relation}
\left(\mathcal{G}_L^{\prime\ast}\right)_{ji}m_j=m_i \left(\mathcal{G}_L^{\prime}\right)_{ji}\quad (i \neq j)\,,
\end{align}
which imply that either $m_i=m_j$ or $\left(\mathcal{G}_L^\prime\right)_{ji}=0$ for $i \neq j$. Thus, $\mathcal{G}_L^\prime$ has to be a real diagonal matrix, if there is no degeneracy in $d_\nu$. This result is completely general and it is just a consequence of the mathematical properties of symmetric matrices~\cite{Lam:2006wm,Grimus:2009pg}. Since neutrino oscillation data requires the light-neutrino mass spectrum to be nondegenerate, we conclude that the maximal symmetry group of $m_\nu$ is indeed $Z_2\times Z_2\times Z_2$. This symmetry is also connected to the freedom~\cite{Branco:2008ai} one has of reversing the ``arrows" of the sides of Majorana unitarity triangles~\cite{AguilarSaavedra:2000vr}.

\section{Symmetries in type I seesaw models and leptogenesis}

In the literature, models that lead to a specific mixing pattern typically impose a symmetry on the Lagrangian and not on the mass matrices. In the most conservative SM extensions, a horizontal symmetry is initially assumed, which is then partially broken so that a residual symmetry remains in the final Lagrangian. Different models and symmetry breaking patterns will then lead to different mass matrix structures.

In the type I seesaw framework, the matrix $m_\nu$ is constructed through the standard seesaw formula
\begin{align}\label{seesaw}
m_\nu=m_DM_R^{-1}m_D^T\,,
\end{align}
which arises from the SM Lagrangian extended with three heavy Majorana neutrino fields $\nu_R$,
\begin{align}\label{lag}
-\mathcal{L}=\overline{\nu_L}m_D\nu_R+\frac{1}{2}\overline{\nu_R^c}M_R\nu_R + \text{H.c.},
\end{align}
after integrating out the heavy fields. Here $m_D$ is the Dirac-neutrino mass matrix, which is a general complex matrix, diagonalizable through unitary matrices $U_{L}^{D}$ and $U_R^{D}$ such that
\begin{align} \label{mDdiag}
U_{L}^{D\dagger}\,m_D\,U_R^{D}=d_D\,,
\end{align}
with $d_D$ a real diagonal matrix. When $m_D$ is Hermitian, $U_{L}^{D}=U_R^{D}$.

Being that $M_R$ is a symmetric matrix, its symmetries can be studied analogously to $m_\nu$. In particular,
\begin{align}\label{GR}
\mathcal{G}_R^TM_R\mathcal{G}_R=M_R\,,
\end{align}
where, as before, the symmetry generators $\mathcal{G}_R$ are built from the columns of the unitary matrix $U_R$ that diagonalizes $M_R$,
\begin{align} \label{MRdiag}
U_R^T M_R U_R=d_R\,,
\end{align}
with $d_R=\text{diag}(M_1,M_2,M_3)$ and $M_i$ real and positive.

Let us next study the implications of a residual symmetry in the Lagrangian of Eq.~(\ref{lag}) for the viability of leptogenesis in flavor models. We recall that the basic quantity in the calculation of the leptogenesis $CP$ asymmetry is the matrix combination $H = m_D^{\prime\dagger} m_D^\prime$, where $m_D^\prime= m_D U_R$ is the Dirac-neutrino mass matrix in the weak basis where the charged lepton and heavy Majorana neutrino mass matrices are diagonal [cf. Eqs.~(\ref{lag}) and (\ref{MRdiag})]. In this basis, the matrix $H$ can be rewritten as
\begin{align} \label{Hmatrix}
H = V_H d_H V^\dagger_H,\quad d_H = |d_D|^2, \quad  V_H=U_R^\dagger U_R^D.
\end{align}
In the context of type I seesaw, this combination appears in the unflavored leptogenesis \textit{CP}-asymmetries $\epsilon_i$ as
\begin{equation} \label{unflavlep}
\epsilon_i \propto \sum_{j \neq i} f_{ij}\, \text{Im}[H_{ij}^2],
\end{equation}
while for the so-called flavored asymmetries $\epsilon_i^\alpha$, the contributions are of the form
\begin{equation} \label{flavlep}
\epsilon_i^\alpha \propto \sum_{j \neq i}\,\text{Im} \{m_{D,i\alpha}^{\prime\dagger} m_{D,\alpha j}^\prime [f_{ij} H_{ij} + g_{ij} H_{ji}]\}\,,
\end{equation}
where $f_{ij}$ and $g_{ij}$ are real scalar functions. In both cases, the off-diagonal elements of the matrix $H$ are the relevant ones, so in order to guarantee a nonzero \textit{CP} asymmetry the following conditions cannot be verified:
\begin{equation}\label{lepcond}
\begin{aligned}
\text{(i)} \quad d_H \propto \openone,\qquad \text{(ii)} \quad V_H = \mathcal{P} K,
\end{aligned}
\end{equation}
where $\mathcal{P}$ are the permutation matrices of three elements, i.e. the set of matrices that represent the $S_3$ group elements, and $K$ is a phase diagonal matrix. If the heavy Majorana neutrino mass spectrum has some degeneracy, then there is an additional freedom to rotate the corresponding sector by a real orthogonal matrix $O$. In this case, $\mathcal{P}$ should be replaced by $O\mathcal{P}$ in condition (ii). An admixture of the two cases is also forbidden.

\subsection{Symmetry of mass matrices as Lagrangian residual symmetry}

In order to have the symmetry of mass matrices as a residual symmetry of the Lagrangian, Eq.~(\ref{lag}) must be invariant under the transformations
\begin{align}\label{fieldtrans}
\nu_L\rightarrow \mathcal{G}_L\nu_L,\quad \nu_R\rightarrow \mathcal{G}_R\nu_R,
\end{align}
where $\mathcal{G}_L$ and $\mathcal{G}_R$ are defined by Eqs.~(\ref{GL}) and (\ref{GR}), respectively. Therefore, from Eq.~(\ref{lag}) we get the additional relation
\begin{align}\label{GLR}
\mathcal{G}_L^\dagger m_D\mathcal{G}_R=m_D\,.
\end{align}

Let us analyze the consequences of this equation for leptogenesis. In the basis where the heavy Majorana neutrinos are diagonal, we can rewrite the symmetry equations as
\begin{align}\label{GH}
\mathcal{G}_R^{^\prime T} d_R\,\mathcal{G}_R^\prime=d_R,\quad \mathcal{G}_R^{\prime \dagger} H\mathcal{G}_R^\prime=H,
\end{align}
with
\begin{equation}\label{GRcons}
\mathcal{G}_R^\prime = U_R^\dagger\mathcal{G}_R U_R.
\end{equation}
Assuming a nondegenerate heavy neutrino mass spectrum, the first relation in Eq.~(\ref{GH}) requires the $Z_2\times Z_2\times Z_2$ symmetry generators $\mathcal{G}_R^\prime$ to be diagonal. Since by construction~\cite{Lam:2006wm,Grimus:2009pg} each symmetry generator $\mathcal{G}_{Ri}\, (i=1,2,3)$ is obtained from a different column of the matrix $U_R$ that diagonalizes $M_R$, the generators $\mathcal{G}_{Ri}^\prime$ cannot take the same form. Their explicit forms are thus given by $\mathcal{G}_{R1}^\prime = \text{diag}(1,1,-1)$, $\mathcal{G}_{R2}^\prime = \text{diag}(1,-1,1)$ and $\mathcal{G}_{R3}^\prime = \text{diag}(-1,1,1)$. The action of any two of these matrices in the second relation of Eq.~(\ref{GH}) would then enforce $H$ to be diagonal. Notice that this conclusion also holds if the symmetry generators $\mathcal{G}_R^\prime$ belong to $SU(3)$ ($\det \mathcal{G}_R^\prime=1$). This in turn implies that one of the conditions given in Eq.~(\ref{lepcond}), or a combination of them, has to be verified, leading then to a vanishing leptogenesis asymmetry.

Clearly, if one does not impose the complete mass matrix symmetry as the residual symmetry of Lagrangian (\ref{lag}), the above conclusions do not necessarily remain valid. For instance, requiring the right-handed sector of the Lagrangian to be invariant just under the transformation $\nu_R \rightarrow \mathcal{G}_{R1}\, \nu_R$ would lead to vanishing $H_{13}$ and $H_{23}$ off-diagonal elements. Yet, a leptogenesis asymmetry could in principle be generated with a nonvanishing $H_{12}$ matrix element.

Let us now consider a degenerate heavy neutrino spectrum. To be specific, let us assume a completely degenerate mass spectrum, i.e. $d_R=M \text{diag}\,(1,1,1)$. The case with double degeneracy trivially follows from this analysis. We start by noticing that in this case $\mathcal{G}_R^{^\prime T}\mathcal{G}_R^\prime=\openone$ and
\begin{align}\label{dH}
\mathcal{G}_R^{\prime\prime\dagger} d_H \mathcal{G}^{\prime\prime}_R=d_H, \quad \mathcal{G}_R^{\prime\prime}=V_H^\dagger\mathcal{G}_R^\prime V_H,
\end{align}
with $d_H$ and $V_H$ defined in Eq.~(\ref{Hmatrix}). Furthermore, if $d_H$ is completely degenerate, then condition (i) in Eq.~(\ref{lepcond}) is verified and the leptogenesis $CP$ asymmetry vanishes. When $d_H$ is partially degenerate, only the nondegenerate sector contributes to leptogenesis. Therefore, we can restrict our analysis to the case when $d_H$ is nondegenerate.

From the orthogonality condition of $\mathcal{G}_R^\prime$ we get
\begin{align}\label{VH}
\mathcal{G}_R^{\prime\prime T}V_H^TV_H\mathcal{G}_R^{\prime\prime}=V_H^TV_H,
\end{align}
which means that $V_H^TV_H$ is diagonal in accordance with Eq.~(\ref{dH}).

Since $V_H$ is a $3\times3$ unitary matrix, it contains nine free parameters. We shall conveniently parametrize it as
\begin{equation} \label{unitpar}
V_H=O_1 K O_2,
\end{equation}
where $O_1$ and $O_2$ are two real orthogonal matrices with three rotation angles each, and $K$ is a phase diagonal matrix with three independent phases. Notice that the above parametrization of a unitary matrix is different from the one commonly used in the literature, which contains three rotation angles and six phases. Using parametrization (\ref{unitpar}) one has
\begin{equation}
V_H^T V_H = O_2^T K^2 O_2.
\end{equation}
In order to have this combination diagonal at least one of the following conditions must be fulfilled:
\begin{equation}
\begin{aligned}
\text{1)} & \quad O_2 =d\,\mathcal{P},\quad d=\text{diag}(\pm 1,\pm 1,\pm 1);\\
\text{2)} & \quad K^2=e^{i\alpha} \openone.
\end{aligned}
\end{equation}
If the first condition is verified, we get from the definition given in Eq.~(\ref{Hmatrix}) that
\begin{equation}
H=O_1\,d_H^\prime\, O_1^T, \quad d_H^\prime = K\,d\,\mathcal{P}\, d_H\,\mathcal{P}^T d\,K^\ast,
\end{equation}
where $d_H^\prime$ is real and diagonal. When the second condition holds,
\begin{equation}
      H=O_1O_2\,d_H\,O_2^TO_1^T.
\end{equation}
In both cases, the matrix $H$ is real and, in general, nondiagonal. Yet, due to the heavy neutrino spectrum degeneracy there is always a freedom to redefine the right-handed fields by an orthogonal transformation, so that all the real off-diagonal entries in $H$ are put to zero and the matrix $H$ is rendered diagonal. Therefore, no leptogenesis \textit{CP} asymmetry can be generated in leading order.

We remark that although two symmetry generators, $\mathcal{G}_L$ and $\mathcal{G}_R$, are introduced in Eq.~(\ref{fieldtrans}), only $\mathcal{G}_R$ is really needed in the above analysis for the proof of vanishing leptogenesis. This is due to the specific form of the matrix combination $H$ that appears in the leptogenesis $CP$ asymmetries.

Before concluding this section let us also note that when $m_D$ is not Hermitian, Eqs.~(\ref{seesaw}) and (\ref{GLR}) imply that the unitary matrices $U_{L}^{D}$ and $U_R^{D}$ that diagonalize $m_D$ have the form
\begin{align}\label{ULR}
U_L^{D}=U_\nu\mathcal{P} K, \quad U_R^{D} =U_R \mathcal{P}^\prime K,
\end{align}
where $\mathcal{P}$ and $\mathcal{P}^\prime$ are two arbitrary permutation matrices. Indeed, from Eq.~(\ref{GLR}) it follows that the matrix $m_D$ is diagonalized on the left by a unitary matrix with the same columns of $U_\nu$ and, on the right, by a unitary matrix with the same columns of $U_R$. This can be easily seen from the construction of the symmetry generators $\mathcal{G}_{L,R}$ through Eqs.~(\ref{GLcons}) and (\ref{GRcons}). Since the order of the columns as well as the addition of any phase diagonal matrix on the right of $U_{\nu}$ and $U_{R}$ are irrelevant in this construction, one has the freedom to redefine the diagonalizing matrices, $U_{\nu,R} \rightarrow U_{\nu,R}\, \mathcal{P}\, K$, which then leads to the form given in Eq.~(\ref{ULR}). Note that the phase diagonal matrix $K$ should be the same in both $U_L^{D}$ and $U_R^{D}$. This follows from the seesaw Eq.~(\ref{seesaw}) and the fact that the diagonal matrix $d_\nu$ defined in Eq.~(\ref{diagm}) is real and positive. On the other hand, the permutation matrices $\mathcal{P}$ and $\mathcal{P}^\prime$ can be different, since they just reflect a reordering of the eigenvalues of $m_D$ and $M_R$. For $m_D$ Hermitian, Eq.~(\ref{GLR}) implies $\mathcal{G}_L=\mathcal{G}_R$ leading to $U_R=U_\nu \mathcal{P} d$. The case when the heavy neutrino sector has some degeneracy can be analogously analyzed since it simply corresponds to the replacement $U_R\rightarrow U_R\,O$, with $O$ a real orthogonal matrix.

We emphasize that the results presented in this section are independent of the mass matrix textures. They are just a consequence of imposing to Lagrangian (\ref{lag}) the maximal residual symmetry, i.e. the symmetry of the mass matrices.

\subsection{Mass-independent textures}

Since a common feature in many flavor models is that the matrices $m_D$, $M_R$, and $m_\nu$ exhibit mass-independent textures, it is worth studying this case and its implications for leptogenesis. In this section we do not look at residual symmetries in the Lagrangian, but rather to the particular textures of the mass matrices in the type I seesaw framework. Let us we rewrite the seesaw formula, Eq.~(\ref{seesaw}), as
\begin{align}\label{dnu}
d_\nu = A\, d_R^{-1} A^T,
\end{align}
where
\begin{equation} \label{Adef}
A = U_\nu^\dagger U_L^{D}\, d_D\, U_R^{D^\dagger} U_R
\end{equation}
is a real matrix, which is independent of the light $m_i$ and heavy $M_i$ neutrino masses.

The above seesaw relation leads to the sets of equations
\begin{align}\label{components}
\sum_k M_k^{-1}\,A^2_{ik}=m_i\,, \quad\sum_{k} M_k^{-1}A_{ik}A_{jk}=0\,,
\end{align}
for the diagonal and off-diagonal elements, respectively. When no degeneracy in the heavy neutrino sector is present, in order to satisfy the second relation of Eqs.~(\ref{components}), at least six of the elements $A_{ij}$ should vanish. If the heavy neutrinos are degenerate, there are additional solutions to these equations which simply correspond to the replacement $U_R \rightarrow U_R\,O$, as allowed by the freedom in rotating the $\nu_R$ fields.

The solutions of Eq.~(\ref{dnu}) or, equivalently, Eqs.~(\ref{components}), can be divided into two classes: $\det\,m_\nu\neq 0$ and $\det\,m_\nu= 0$, which we discuss next in detail.

\subsubsection{$\det\,m_\nu \neq 0$}

Since the matrix $A$ has at least six zero entries and $\det\,A = \det\,d_D \neq 0$, its possible textures are of the form of a permutation matrix. This in turn implies that the matrices $A A^\dagger$ and $A^\dagger A$ should be diagonal. From the definition of $A$ given in Eq.~(\ref{Adef}), we then conclude that
\begin{equation}
U_\nu^\dagger U_L^D = \mathcal{P} K, \quad  U_R^{D\dagger} U_R = K^\ast \mathcal{P}^\prime,
\end{equation}
which are equivalent to the relations given in Eq.~(\ref{ULR}). These relations clearly reflect  the required correlations between the different unitary matrices that diagonalize $m_D$, $M_R$, and $m_\nu$ with mass-independent textures and, simultaneously, verify the seesaw formula (\ref{dnu}).

Thus, if $\det\,m_\nu \neq 0$, the assumption of mass-independent textures leads to the symmetry relation given in Eq.~(\ref{GLR}). In other words, Lagrangian (\ref{lag}) exhibits the maximal residual symmetry, i.e. the symmetry of mass matrices. We remark that this result holds for any mass-independent texture model, and, in particular, for flavor models that predict a mass-independent leptonic mixing. The conclusions for leptogenesis are therefore exactly the same as before, namely, the vanishing of the leptogenesis asymmetry in leading order.

\subsubsection{$\det\,m_\nu = 0$}

The present neutrino oscillation data does not preclude the existence of a massless neutrino, so it is pertinent to analyze this case. Since in the relevant basis for leptogenesis $m_D^\prime=U_\nu A$, the combination $H=m_D^{\prime\dagger} m_D^\prime=A^TA$ is always real, thus forbidding unflavored leptogenesis. From the first relation in Eqs.~(\ref{components}) one sees that the $i$-th row of $A$ corresponding to $m_i=0$ is null. Moreover, the second relation prohibits the coexistence of two nonvanishing elements in any column of $A$.  Considering the very restrictive set of $A$-textures that verify Eqs.~(\ref{components}), it is easy to show that the combination $m_{D,\alpha i}^{\prime\ast} m_{D,\alpha j}^\prime=\sum_{k,k^\prime}U^\ast_{\alpha k} U_{\alpha k^\prime} A_{k i}A_{k^\prime j}$ always vanishes unless $k=k^\prime$. In the latter case, the above combination is always real, which then implies that the $CP$ asymmetries given in Eq.~(\ref{flavlep}) are equal to zero and flavored leptogenesis is also forbidden.

\section{Conclusions}

In conclusion, we have shown that the type I seesaw flavor models in the literature that lead to a mass-independent leptonic mixing do not have leptogenesis in leading order. In particular, in models that predict the HPS tribimaximal mixing, the residual symmetry of the Lagrangian is the symmetry of the mass matrices and a symmetry relation connecting the left and right sectors through the Dirac-neutrino mass matrix is obtained. As a consequence, in these models the Dirac-neutrino Yukawa coupling combinations relevant for leptogenesis can be rendered diagonal in the weak basis where the charged leptons and heavy Majorana neutrinos are diagonal.

Finally we remark that our conclusions do not necessarily remain valid if other types of seesaw mechanisms are invoked. For instance, for a hybrid type I/II seesaw, where a scalar $SU(2)_L$ triplet is added, combinations such as $\text{Im}[(m^\dagger_D m_\nu^{II} m_D^\ast)_{kk}]$ and $\text{Im}[\text{Tr}[m_\nu^{II \ast}m_\nu^{I}]]$, where $m_\nu^{I,II}$ are the type I and type II seesaw contributions to the effective neutrino mass matrix $m_\nu$, could also appear in the leptogenesis \textit{CP} asymmetries~\cite{Hambye:2003ka}. It can be easily shown that hybrid type I/II seesaw flavor models belonging to any of the cases studied here give, in general, no restrictions for leptogenesis.

\emph{Note added in proof: } After the completion of this work, Refs.~\cite{Bertuzzo:2009im,AristizabalSierra:2009ex} have appeared where some of the issues considered here have been partially addressed.

\begin{acknowledgments}
We are very grateful to G.C. Branco for invaluable discussions and useful comments. The work of H.S. was supported by {\em Funda\c{c}\~{a}o para a Ci\^{e}ncia e a Tecnologia} (FCT, Portugal) under the Grant No. SFRH/BD/36994/2007. This work was partially supported by FCT through the projects CFTP-FCT UNIT 777 and CERN/FP/83502/2008, which are partially funded through POCTI (FEDER), and by the Marie Curie RTNs MRT-CT-2006-035505.
\end{acknowledgments}

  \end{document}